\documentclass[nocite, overfull]{epl}

\usepackage{amssymb}

\title{Anomalous Nernst Effect in the Vortex-Liquid Phase of
High-Temperature Superconductors by Layer Decoupling}
\author{J.P. Rodriguez\inst{1}}

\institute{                    
  \inst{1} Department of Physics and Astronomy, California State University, Los Angeles, CA 90032,
USA}
\pacs{74.25.Dw}{First pacs description}
\pacs{74.25.Ha}{Second pacs description}
\pacs{74.25.Qt}{Third pacs description}

\begin{document}

\maketitle

\begin{abstract}
Linear diamagnetism is predicted in the vortex-liquid phase of layered superconductors
at temperatures just below the mean-field phase transition on the basis of
a high-temperature analysis of the corresponding frustrated $XY$ model.  
The diamagnetic susceptibility, and the Nernst signal by implication,
is found to vanish with temperature as
$(T_{c0} - T)^3$ in the vicinity of the meanfield transition at $T_{c0}$.
Quantitative agreement with recent experimental observations of a diamagnetic signal
in the vortex-liquid phase of high-temperature superconductors is obtained.
\end{abstract}

\section{Introduction}
The Abrikosov vortex lattice melts into an extended vortex-liquid phase
in high-temperature superconductors
subject to an
external magnetic field oriented perpendicular to the conducting copper-oxygen planes 
that make them up\cite{zeldov06}\cite{crabtree00}.
The large size in temperature and magnetic field 
of the vortex-liquid phase can
be attributed to such layer anisotropy\cite{g-k91}\cite{daemen}\cite{jpr00}.
A cross-over from a vortex-line liquid at temperatures
just above the melting point of the Abrikosov vortex lattice to
a decoupled vortex liquid at higher temperature
that shows negligible correlations
of the superconducting order parameter across layers
is predicted if the vortex lattice in isolated layers melts through
a continuous or a weakly first-order phase transition\cite{jpr02}.
Such dimensional cross-over is observed experimentally in
electronic transport studies of
the vortex-liquid phase in moderately anisotropic  
high-temperature superconductors\cite{qiu00}.
The Abrikosov vortex lattice is predicted to sublimate directly into a decoupled vortex
liquid at large enough layer anisotropy, on the other hand,
if the vortex lattice in isolated layers melts through a first-order
phase transition\cite{jpr02}.
Electronic transport studies of the mixed phase
in extremely layered high-temperature superconductors are consistent with
the last sublimation scenario\cite{fuchs97}.

An anomalous Nernst effect is also observed in the vortex-liquid phase of high-temperature
superconductors\cite{ong06}.
In particular, a gradient in temperature along  the copper-oxygen planes
generates an electric field perpendicular to it
along the copper-oxygen planes as well.
The low-temperature onset of the anomalous Nernst signal coincides with the
melting point of the Abrikosov vortex lattice,
while the high-temperature onset can lie above the critical temperature
of the superconducting state at zero field.
The authors of ref. \cite{ong06} argue that this 
effect is principally due to vortex excitations in the mixed phase
of high-temperature superconductors.
It is then tempting to identify the cross-over between three-dimensional (3D)
and two-dimensional (2D) vortex-liquid behavior
that is 
predicted for layered superconductors in certain instances\cite{jpr02}
with the peak in the Nernst signal.
The fact that anomalous Nernst signals are also observed in the vortex-liquid
phase of extremely layered high-temperature superconductors 
that do not show the former dimensional cross-over\cite{fuchs97}\cite{ong06}
rules out that interpretation, however.

The anomalous Nernst effect 
observed in the vortex-liquid phase of high-temperature superconductors
may instead be principally due to vortex excitations
in copper-oxygen planes
that are virtually isolated from one another\cite{ong06}.
In this Letter,
the theoretical consequences of that proposal
are examined
through a duality analysis of the uniformly frustrated $XY$ model for the
mixed phase of extremely type-II superconductors\cite{jpr00}\cite{jpr02}.
We find first that weak collective pinning of the vortex lattice
results in a melting/decoupling temperature
that does {\it not} extrapolate to the mean-field transition in zero field.  
Instead,
a relatively big region of vortex liquid  that is stabilized by random pinning centers 
is predicted to exist at temperatures below 
the mean-field transition.
Second, a high-temperature expansion of the uniformly frustrated $XY$ model
yields linear diamagnetism at temperatures just below the mean-field transition.
The temperature dependence of the predicted equilibrium magnetization
is found to agree quantitatively with recent experimental reports of a diamagnetic
signal extracted from the vortex-liquid phase of high-temperature superconductors\cite{ong05b}.
Last, 
we emphasize that an anomalous Nernst effect is generally expected
inside of the vortex liquid phase\cite{ong06},
where it tracks the temperature dependence shown by the diamagnetism
in the vicinity of the mean-field phase transition.

\section{Vortex-Lattice Melting/Decoupling}
The $XY$ model with uniform frustration is the minimum
theoretical description of  vortex matter in
extremely type-II 
superconductors.  Both fluctuations of the magnetic induction and
of the magnitude of the superconducting order parameter are neglected
within this approximation.
The model hence  is valid deep inside
the interior of the mixed phase.  
Its thermodynamics
is determined by the superfluid kinetic energy
\begin{equation}
E_{XY}^{(3)} = -\sum_{r}\sum_{\mu=x,y,z} J_{\mu} {\rm cos}
[\Delta_{\mu}\phi  - A_{\mu}]|_{r},
\label{3dxy}
\end{equation}
which is a  functional of the phase 
of the superconducting order parameter,
$e^{i \phi}$,
over the cubic lattice, $r$.
Here, $J_x$ and   $J_y$ 
denote the local phase rigidities over nearest-neighbor links  within layers.
These are equal and constant,
except over links in the vicinity
of a pinning center.
The Josephson coupling across adjacent layers, $J_z$,
shall be assumed to be constant and weak.
It can be parameterized by
$J_z = J_0 /\gamma^{\prime 2}$, 
where $J_0$ is the Gaussian stiffness of the $XY$ model for each layer in isolation,
and where $\gamma^{\prime}$ is the model anisotropy parameter.
The   vector potential
$A_{\mu} = (0, 2\pi f x/a, 0)$
represents the magnetic induction
oriented perpendicular to the layers,
$B_{\perp} = \Phi_0 f / a^2$.
Here $a$ denotes the square lattice constant, which is of order
the coherence length of the Cooper pairs, $\Phi_0$ denotes
the flux quantum, and $f$ denotes the concentration of vortices per site.

The thermal/bulk average of the Josephson coupling between adjacent layers
is given by
the expression\cite{jpr00}\cite{koshelev96}
\begin{equation}
\overline{\langle {\rm cos}\, \phi_{l, l+1}\rangle} \cong 
y_0 \sum_1 \overline{C_l (0,1) \cdot C_{l+1}^* (0,1)}
e^{i[A_z (1) - A_z (0)]}
\label{cos1}
\end{equation}
in the decoupled vortex liquid
to lowest order in the fugacity
$y_0 = J_z / 2 k_B T$.
Here
$\phi_{l, l+1} (\vec r) =  
\phi (\vec r, l + 1) - \phi (\vec r, l) - A_z (\vec r)$
is the gauge-invariant phase difference across adjacent layers $l$ and $l+1$,
and
$C_l (1, 2) = \langle e^{-i\phi(1)} e^{i\phi(2)}\rangle_0$
is the autocorrelation function of the superconducting order parameter
within layer $l$ in isolation ($J_z = 0$).
Short-range  correlations
on the scale of  $\xi_{2D}$
following
$C_l (1,2) = g_0 e^{-r_{1,2}/\xi_{2D}} 
e^{-i\phi_0 (1)} e^{i\phi_0 (2)}$
yields the result\cite{jpr04}
\begin{equation}
\overline{\langle {\rm cos}\, \phi_{l, l+1}\rangle} \sim
g_0^2 (J_0 / k_B T)   [(l_{\phi}^{-1} + \xi_{\phi}^{-1})^{-1} / \Lambda_0]^2
\label{cos2}
\end{equation}
for the inter-layer ``cosine'' (\ref{cos1}).
Here, $l_{\phi}$ is a quenched disorder scale for the vortex lattices 
{\it across} adjacent pairs of isolated layers 
that appears through the autocorrelation
\begin{equation}
\overline{{\rm exp} [{i\phi_{l, l + 1}^{(0)} (1)}]
\cdot 
{\rm exp} [{-i\phi_{l, l + 1}^{(0)} (2)}]}
=
 e^{-r_{1,2}/l_{\phi}}.
\label{form3}
\end{equation}
of the quenched inter-layer phase difference, 
$\phi_{l, l+1}^{(0)} (\vec r) = 
\phi_0 (\vec r,  l + 1)
 - \phi_0 (\vec r,  l) - A_z (\vec r)$. 
It is set by the density of dislocations quenched into the 2D vortex lattices
found in each layer at zero temperature
in the present case of uncorrelated pinning centers.
Also, above we have
$\xi_{\phi} = \xi_{2D} / 2$ and
the Josephson penetration depth
$\Lambda_0 = \gamma^{\prime} a$. 

\begin{figure}
\onefigure[scale=0.31, angle=-90]{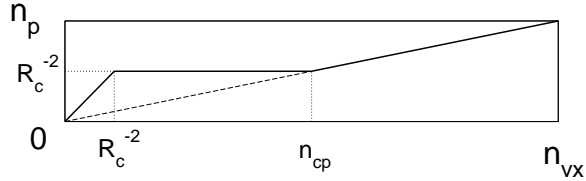}
\caption{Schematic profile of the density of pinned vortices versus the total density
of vortices
within an isolated layer.  Pinning centers are assumed {\it not} to crowd together.}
\label{f.1}
\end{figure}

In the absence of inter-layer coupling, 
arbitrarily weak random point pins result in a 
stack of 2D vortex lattices with dislocations quenched in\cite{shi91}.
Let us assume that each 2D vortex lattice is in a {\it hexatic vortex glass} state\cite{jpr05},
such that dislocations do {\it not} arrange themselves into grain boundaries.
The quenched disorder scale $l_{\phi}$ that renormalizes down the interlayer Josephson coupling
(\ref{cos2}) is then set by the density of such dislocations\cite{jpr04}.
Recent theoretical calculations find that
each isolated layer shows a net superfluid density near zero temperature
in the collective pinning regime,
where the number of dislocations quenched into each 2D vortex lattice is small in comparison to the number
of pinned vortices\cite{jpr05}.
Application of collective pinning theory to the 2D vortex lattices found in isolated layers
yields a density of quenched-in dislocations
identical to the density of Larkin domains\cite{m-e}
$R_c^{-2} \sim n_{\rm p} (f_{\rm p} / \nu_0 b)^2$,
where $n_{\rm p}$ denotes the density of pinned vortices per layer,
where $f_{\rm p}$ denotes the maximum pinning force, 
and where $\nu_0$ denotes the shear modulus of the 2D vortex lattice.
Here the critical state is assumed to be limited by
plastic creep of Larkin domains 
by an elementary burgers vector $\vec b$ of the 2D vortex lattice.
Consider now
the limit of weak pinning centers
that do not crowd together:
$f_{\rm p}\rightarrow 0$ and $\pi r_{\rm p}^2\cdot n_0 \ll 1$, respectively,
where $n_{0}$ denotes the density of pinning centers per layer,
and where $r_{\rm p}$ denotes the range of each  pinning center.
Simple probabilistic considerations then yield the
identity $n_{\rm p}/ n_{0} = \pi r_{\rm p}^2 / a_{\rm vx}^2$
between the fraction of occupied pinning centers and the ratio of
the effective area of each pinning center to the area per vortex, $a_{\rm vx}^2 = a^2 / f$.
This yields the result $n_{\rm p} = (n_{0}\cdot \pi r_{\rm p}^2) n_{\rm vx}$
for the density of pinned vortices\cite{jpr-maley}, 
where $n_{\rm vx} = 1/a_{\rm vx}^2$ is the density of vortices per layer.
Finally,
substitution of the estimate $\nu_0 = (\pi / 4) n_{\rm vx} J_0$ for the shear modulus\cite{brandt}
yields the result 
$R_c^{-2} \sim (f_{\rm p} r_{\rm p} / J_0)^2 n_{0}$ 
for the density of Larkin domains,
which is
independent of magnetic field.  
Note, however, that all of the above is valid only
in the 2D collective pinning regime that exists
at perpendicular magnetic fields above the threshold
$B_{\rm cp}^{(2D)} \sim (f_{\rm p}/J_0)^2 \Phi_0$,
in which case many vortices are pinned in each layer
within a Larkin domain of dimensions $R_c\times R_c$ \cite{jpr05}.
Single-vortex pinning exists at magnetic field below that threshold,
on the other hand,
in which case each Larkin domain contains only a single pinned vortex: 
$n_{\rm p} = R_c^{-2}$.
Assembling the above suggests the profile for the density of pinned vortices per layer
versus the density of vortices 
that is depicted by fig.~\ref{f.1}.
It implies that the quenched disorder scale $l_{\phi}\sim R_c$
is independent of magnetic fields above the threshold
$B_{L}^{(2D)} \sim \Phi_0 / R_c^2$.

\begin{figure}
\onefigure[scale=0.31, angle=-90]{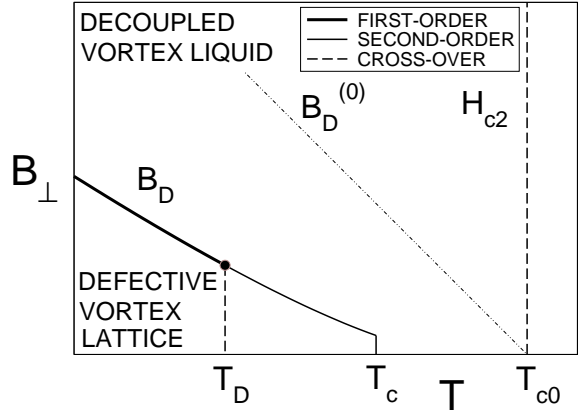}
\caption{Schematic phase diagram near the critical temperature at zero magnetic field.
Continuity with the zero-field limit of the $XY$ model (\ref{3dxy})
implies a {\it lower} critical point, which is observed experimentally in the
phase diagram of high-temperature superconductors with point disorder. (See ref. \cite{crabtree00}.)}
\label{f.2}
\end{figure}

We are finally in a position to determine the melting/decoupling line of the 3D vortex lattice
at temperatures outside of the 2D critical regime, $\xi_{2D}\sim a_{\rm vx}$,
at big enough perpendicular magnetic fields
such that Larkin domains can be defined,
$B_{\perp} \gg B_{L}^{(2D)}$.
The identification of the separation between dislocations quenched into each 2D vortex lattice 
with the 2D Larkin scale\cite{m-e},
$l_{\phi}\sim R_c$,
necessarily yields the inequality 
$\xi_{\phi} \ll l_{\phi}$ in such case.
At temperatures lying
inside of the interval $[T_m^{(2D)}, T_c^{(2D)}]$
bounded by melting  of the  2D vortex-lattice 
and by the Kosterlitz-Thouless transition in isolated layers,
yet lying outside of the 2D critical regime,
a partial duality analysis of the pristine layered $XY$ model with uniform frustration (\ref{3dxy})
finds a first-order melting/decoupling transition of the 3D vortex lattice 
at interlayer Josephson coupling\cite{jpr00}
\begin{equation}
\overline{\langle {\rm cos}\, \phi_{l, l+1}\rangle} \simeq 1/2.
\label{xover}
\end{equation}
The first-order nature of this melting/decoupling line
and its coincidence with  the contour defined above is
consistent both with Monte Carlo simulations of the same model\cite{koshelev96}
and with elastic medium descriptions of the vortex lattice in layered
superconductors\cite{daemen}.
Observe now that the criterion (\ref{xover}) for first-order melting/decoupling
should remain valid in the present  regime of weak pinning such that $l_{\phi} \gg \xi_{\phi}$.
Substitution of expression (\ref{cos2}) for the inter-layer ``cosine''
in the decoupled vortex liquid then yields the melting/decoupling field
\begin{equation}
B_D = \Bigl(\sqrt{B_D^{(0)}} - \sqrt{B_{L}^{(2D)}}\Bigr)^2,
\label{b_d}
\end{equation}
where $B_D^{(0)} \sim g_0^2 (J_0 / k_B T) (\Phi_0 / \Lambda_0^2)$
is the melting/decoupling field in the pristine limit\cite{daemen}\cite{jpr00}\cite{koshelev96}.
These results are summarized by the phase diagram shown in fig.~\ref{f.2}.
The short sections of dashed and solid lines that emanate perpendicularly from the horizontal axis
originate respectively
from the decoupling cross-over
(\ref{xover}) and the second-order phase transition 
shown by the layered $XY$ model in the absence of uniform frustration (cf. ref. \cite{g-k90}). 
We conclude this section by observing that
the melting/decoupling line does {\it not} extrapolate to the mean-field critical temperature
at zero-field [$J_0 (T_{c0}) = 0$] 
due to the presence of dislocations quenched into
the weakly pinned vortex lattices found in isolated layers.

\section{Vortex-Liquid Diamagnetism}  The phase diagram for the mixed phase of layered
superconductors shown by fig.~\ref{f.2} implies a large region of vortex liquid
in the vicinity of the meanfield transition at zero magnetic field
because of the effects of random point pins.  In particular, the equilibrium diamagnetic
susceptibility due to the emergence of Cooper pairs
is well
defined at temperatures inside of the window $[T_c, T_{c0}]$.
The former quantity can be obtained from the uniformly frustrated $XY$ model (\ref{3dxy})
in the vicinity of the meanfield transition via a high-temperature expansion
in powers of the fugacity $z_0 = J  / 2 k_B T$ \cite{drouffe}.  
In particular, a duality analysis yields that
the corresponding partition function is approximated by
$Z_{XY} \cong Z_0 + Z_4$ as $z_0\rightarrow 0$, 
where 
$Z_0 = \prod_{\langle ij\rangle} I_0 (J_{ij}/ k_B T)$,
and where
$Z_4 / Z_0 = 2\,{\rm cos} (2\pi f) \sum_{\Box} \prod_{\langle ij\rangle\in\Box} t_1 (J_{ij}/ k_B T)$.
Here $t_1 (x) = I_1 (x) / I_0 (x)$ is the ratio between  a first-order and a
zero-order modified Bessel function, 
which is approximately $t_1 (x) \cong x/2$ for $|x|\ll 1$.
Also, $\langle ij\rangle$ represents nearest-neighbor links
within layers, and $\Box$ represents elementary plaquettes within layers.
The equilibrium magnetization is given by
$M_{\perp} = - \partial (G_{XY}/V) / \partial B_{\perp}$
in the extreme type-II limit, 
where
$G_{XY} = - k_B T \cdot {\rm ln}\, Z_{XY}$
is the Gibbs free energy, and where $V$ is the volume.
Substitution of the previous high-temperature approximation yields
\begin{equation}
M_{\perp} = - (2\pi / \Phi_0) (J_{4} / d) (J_{4} / 2 k_B T)^3 \, {\rm sin} (2\pi f),
\label{M}
\end{equation}
where $J_{4} = (\overline{\prod_{\langle ij\rangle\in\Box} J_{ij}})^{1/4}$,
and where $d$ denotes the spacing between layers\cite{artifact}.
The magnetization therefore varies linearly with vanishing magnetic field like
$M_{\perp} =  \chi H_{\perp}$,
with a diamagnetic susceptibility
\begin{equation}
4\pi \chi = -\kappa^{-2} (J_0 / 2 k_B T)^3 (J_{4}/J_0)^4 (a/\xi)^2.
\label{chi1}
\end{equation}
Here $\kappa = \lambda_L / \xi$ is the usual ratio
of the London penetration depth
to the coherence length of the Cooper pairs.
The former is related to the Gaussian phase stiffness of each layer by
$J_0 = \Phi_0^2 d/16\pi^3 \lambda_L^2$.

\begin{figure}
\onefigure[scale=0.31, angle=-90]{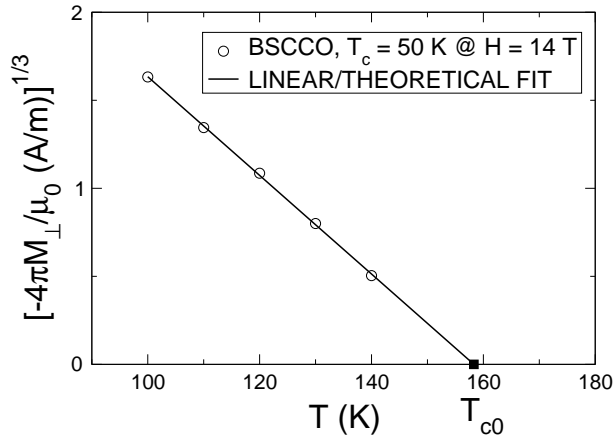}
\caption{Fit of diamagnetic signal extracted from underdoped BSCCO (ref. \cite{ong05b})
to the theoretical prediction near the mean-field transition [Eq. (\ref{chi2})].}
\label{f.3}
\end{figure}

Non-linear diamagnetism is observed experimentally in the vortex liquid phase of
the extremely layered high-temperature superconductor Bi$_2$Sr$_2$CaCu$_2$O$_{8+\delta}$ (BSCCO),
at temperatures just above the  superconducting transition in zero field\cite{ong05a}.
Linear diamagnetism is displayed
at yet higher temperature in the same samples, on the other hand.
By Eq. (\ref{M}),
the uniformly frustrated $XY$ model (\ref{3dxy})
predicts such linear diamagnetism, $M_{\perp} = \chi H_{\perp}$, 
at perpendicular magnetic fields that are small compared to the upper-critical scale
$\Phi_0 / 2 \pi a^2$,
at temperatures just below the mean-field phase transition.
The corresponding diamagnetic susceptibility predicted by the $XY$ model is given by
Eq. (\ref{chi1}). 
Use of the relation quoted previously between the Gaussian phase rigidity within planes
and the London penetration length
in conjunction with physical parameters
$\kappa = 100$, $\lambda_L (0) = 0.2 \, \mu{\rm m}$, and $d = 1.5$ nm 
appropriate for BSCCO\cite{Tinkham} 
yields the estimate
\begin{equation}
4\pi \chi / \mu_0 = - [(503 \, ^{\circ} {\rm K}  / T) (n_s / n)]^3 (J_{4}/J_0)^4 (a/\xi)^2 \, {\rm A/Tm}
\label{chi2}
\end{equation}
for the diamagnetic susceptibility of that material in the vicinity of the mean-field phase transition.
Here $n_s (T) / n = \lambda_L^2 (0) / \lambda_L^2 (T)$ is the superfluid fraction.
The mean-field superfluid density expected from 
a pristine $d$-wave state in 2D is approximately $2/3$ the corresponding $s$-wave result in the
vicinity of the meanfield transition at zero field\cite{maki94}; 
i.e., $n_s/n \cong (4/3)(1-t)$, where  $t = T/T_{c0}$
is the reduced temperature.
Equation (\ref{chi2}) then implies that the diamagnetic susceptibility
vanishes like $(1 - t)^3$ with temperature
as it approaches the mean-field transition. 
Figure~\ref{f.3} displays the cube-root of the diamagnetic signal extracted experimentally
in ref. \cite{ong05b} from an underdoped sample of BSCCO
with $T_c = 50^{\circ}$ K, 
in perpendicular magnetic field $H_{\perp} = 14$ T,
as a function of temperature.  
The solid line is a fit to the linear diamagnetism,
$M_{\perp} = \chi H_{\perp}$, predicted by
the  high-temperature expansion of the uniformly frustrated $XY$ model,
Eq. (\ref{chi2}),
with $n_s/n = (4/3)t(1-t)$, $T_{c0} = 158^{\circ}$ K, 
and with $XY$ model parameter $a = 0.28\,(J_0 / J_4)^2 \xi$.
The success of the fit indicates that the onset of the diamagnetic
signal observed in the vortex liquid phase of high-temperature
superconductors reflects nothing other than the mean-field phase transition
at which Cooper pairs emerge.
The large suppression of $T_c$ compared to the meanfield transition temperature $T_{c0}$
obtained here can be accounted for by quenched  disordering
of the superconducting order parameter,
which could be generic to under-doped high-temperature superconductors\cite{emery-kivelson}.

\section{Anomalous Nernst Effect and Conclusions}
A gradient in temperature along the layers in the
vortex liquid phase of high-temperature superconductors 
generates a voltage in the perpendicular direction
within the layers\cite{ong06}.  In particular, 
the Nernst signal
defined by the ratio $e_y = E_y / \partial_x T$
between the electric field that is generated and the gradient in temperature
peaks inside of the vortex liquid.
Standard transport theory yields the identity\cite{ong06}
\begin{equation}
e_y = \rho_x \cdot \alpha_{xy}^{s}
\label{e_y}
\end{equation}
between the Nernst signal and the product
of the flux-flow electrical resistivity $\rho_x$
with the off-diagonal Peltier coefficient $\alpha_{xy}^s$.
Also,
application of
Ginzburg-Landau theory for the superconducting order parameter
yields the estimate\cite{maki67} 
\begin{equation}
\alpha_{xy}^{s} \cong \bar\beta M_{\perp}
\label{alpha}
\end{equation}
for the Peltier coefficient
near the mean-field transition,
where $\bar\beta$ is of order $T^{-1}$.
Observe now that the flux-flow resistance increases with temperature in the vortex liquid,
while the equilibrium magnetization decreases with temperature there [cf. Eq. (\ref{chi1})].
Substitution of the estimate (\ref{alpha})
into the identity (\ref{e_y}) then yields
({\it i}) that the low-temperature onset of the anomalous Nernst signal
is given by the melting/decoupling temperature of the vortex lattice.
Also, 
the linear diamagnetism (\ref{chi1}) extracted from the high-temperature regime of
the frustrated $XY$ model
implies 
({\it ii}) that the anomalous Nernst
signal vanishes with temperature at the mean-field transition as $(T_{c0} - T)^3$.
Where exactly the Nernst signal peaks inside of the vortex-liquid phase
depends on how pinning affects the flux-flow resistance\cite{Tinkham},
which is beyond the scope of the paper.

In conclusion, a high-temperature analysis of the layered $XY$ model with uniform
frustration finds that the simultaneous onset of linear diamagnetism and 
of an
anomalous Nernst effect in 
the normal phase of
high-temperature superconductors\cite{ong05b}
can be identified with the mean-field transition for Cooper pairing.
The low-temperature onset of the anomalous Nernst signal
at the melting/decoupling line of the vortex lattice
was also found to be depressed substantially by the presence of 
dislocations quenched into the vortex lattice in isolated layers.

\acknowledgments
The author thanks Louis Taillefer for discussions
and P.W. Anderson for correspondence.
This work was supported in part by the US Air Force
Office of Scientific Research under grant no. FA9550-06-1-0479.

\end{document}